\documentclass[acmsmall,screen]{acmart}

\usepackage{algorithm}
\usepackage[noend]{algpseudocode}
\usepackage{multirow}
\usepackage{booktabs} 
\usepackage{amsmath}
\usepackage{graphicx}
\usepackage{caption}
\usepackage[T1]{fontenc}
\usepackage[utf8]{inputenc} %
\usepackage{microtype}

\usepackage{subcaption}
\usepackage{rotating}
\usepackage{adjustbox}
\usepackage{array}
\usepackage{float}
\usepackage{tabularx}
\usepackage{xcolor}
\usepackage{stfloats}
\usepackage{wrapfig}
\usepackage{listings}
\usepackage{color, colortbl}
\usepackage{balance} 
\usepackage{lscape}
\usepackage{xspace}
\usepackage{framed}
\usepackage{syntax}
\usepackage{parcolumns}
\usepackage{pifont}
\usepackage{soul}
\usepackage[tikz]{bclogo}
\usepackage{enumitem}
\usepackage{soul}
\usepackage{makecell}
\usepackage{tcolorbox}
\usepackage{enumitem}

\usepackage{amssymb}
\usepackage{wasysym}

\newcommand{\boxmargin}{1mm}
\newtcolorbox{myboxc}{
    colback=gray!15!white,
    arc = 0pt, outer arc = 0pt,
    boxsep=0pt, left = 3pt, right = 0pt, top = 0pt, bottom = 0pt, 
    leftrule=3pt, bottomrule=0pt,toprule=0pt, rightrule=0pt,
    left = \boxmargin, right = \boxmargin, top = \boxmargin, bottom = \boxmargin
}

\newcommand{\Swebenchx}{OmniGIRL\xspace}
\newcommand{\Claude}{Claude-3.5-Sonnet\xspace}
\newcommand{\GPT}{GPT-4o\xspace}
\newcommand{\DeepSeek}{DeepSeek-V2.5\xspace}
\newcommand{\ClaudeFullName}{Claude-3.5-Sonnet-2024-06-25\xspace}
\newcommand{\GPTFullName}{GPT-4o-2024-08-06\xspace}
\newcommand{\DeepSeekFullName}{DeepSeek-V2.5\xspace}
\newcommand{\secmargin}{\vspace{-3mm}} 
\newcommand{\figmargin}{\vspace{-3mm}} 
\newcommand{\tabmargin}{\vspace{-2mm}} 

\AtBeginDocument{%
  }

\setcopyright{cc}
\setcctype{by-nc-sa}
\acmDOI{10.1145/3728871}
\acmYear{2025}
\acmJournal{PACMSE}
\acmVolume{2}
\acmNumber{ISSTA}
\acmArticle{ISSTA002}
\acmMonth{7}
\received{2024-10-31}
\received[accepted]{2025-03-31}

\begin{document}

\title{\Swebenchx: A Multilingual and Multimodal Benchmark for GitHub Issue Resolution}

\author{Lianghong Guo}
\orcid{0009-0001-0943-5049}
\affiliation{
  \institution{Sun Yat-sen University, Zhuhai Key Laboratory of Trusted Large Language Models}
  \city{Zhuhai}
  \country{China}
}
\email{guolh8@mail2.sysu.edu.cn}

\author{Wei Tao}
\orcid{0000-0002-1800-1904}
\affiliation{
  \institution{Independent Researcher}
  \city{Shenzhen}
  \country{China}
}
\email{wtao@ieee.org}

\author{Runhan Jiang}
\orcid{0009-0000-3343-8023}
\affiliation{
  \institution{Sun Yat-sen University}
  \city{Zhuhai}
  \country{China}
}
\email{guolh8@mail2.sysu.edu.cn}

\author{Yanlin Wang}
\authornote{Corresponding author.}
\orcid{0000-0001-7761-7269}
\affiliation{
  \institution{Sun Yat-sen University, Zhuhai Key Laboratory of Trusted Large Language Models}
  \city{Zhuhai}
  \country{China}
}
\email{wangylin36@mail.sysu.edu.cn}

\author{Jiachi Chen}
\orcid{0000-0002-0192-9992}
\affiliation{
  \institution{Sun Yat-sen University, Zhuhai Key Laboratory of Trusted Large Language Models}
  \city{Zhuhai}
  \country{China}
}
\email{chenjch86@mail.sysu.edu.cn}

\author{Xilin Liu}
\orcid{0009-0001-4870-1012}
\affiliation{
  \institution{Huawei Cloud Computing Technologies Co., Ltd.}
  \city{Shenzhen}
  \country{China}
}
\email{liuxilin3@huawei.com}

\author{Yuchi Ma}
\orcid{0009-0002-3304-1389}
\affiliation{
  \institution{Huawei Cloud Computing Technologies Co., Ltd.}
  \city{Shenzhen}
  \country{China}
}
\email{mayuchi1@huawei.com}

\author{Mingzhi Mao}
\orcid{0000-0001-9369-7828}
\affiliation{
  \institution{Sun Yat-sen University}
  \city{Zhuhai}
  \country{China}
}
\email{mcsmmz@mail.sysu.edu.cn}

\author{Hongyu Zhang}
\orcid{0000-0002-3063-9425}
\affiliation{
  \institution{Chongqing University}
  \city{Chongqing}
  \country{China}
}
\email{hyzhang@cqu.edu.cn}

\author{Zibin Zheng}
\orcid{0000-0002-7878-4330}
\affiliation{
  \institution{Sun Yat-sen University, Zhuhai Key Laboratory of Trusted Large Language Models}
  \city{Zhuhai}
  \country{China}
}
\email{zhzibin@mail.sysu.edu.cn}


\begin{teaserfigure}
\fontsize{12}{6}\selectfont
\begin{tabular}{@{\hspace{2em}}l@{\hspace{6.5em}}c@{\hspace{5.5em}}r@{}}
  \href{https://deepsoftwareanalytics.github.io/omnigirl_leaderboard.html}{%
    \includegraphics[height=1.8ex]{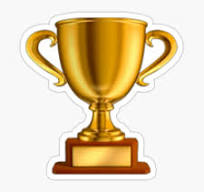}\,\textmd{Leaderboard}} &
  \href{https://huggingface.co/datasets/Deep-Software-Analytics/OmniGIRL}{%
    \includegraphics[height=1.8ex]{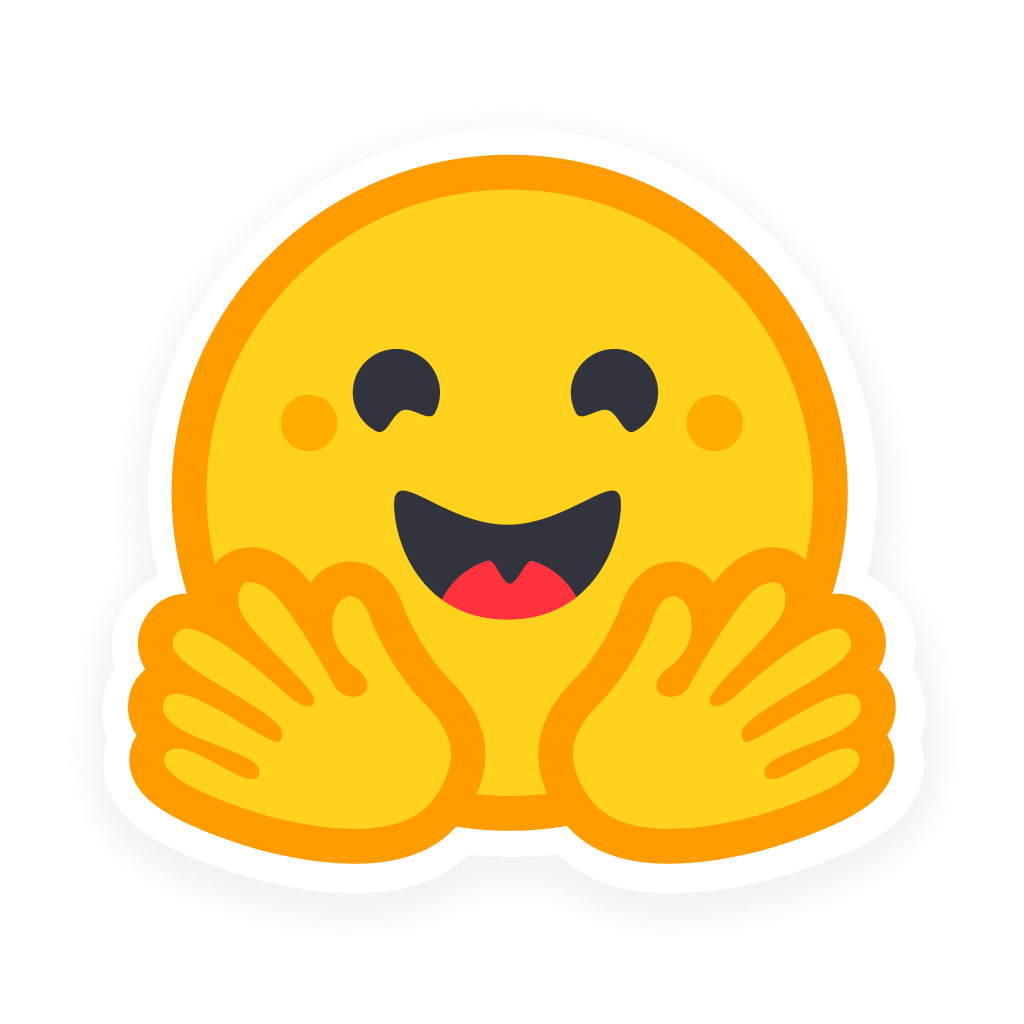}\,\textmd{Dataset}} &
  \href{https://github.com/DeepSoftwareAnalytics/OmniGIRL}{%
    \includegraphics[height=1.8ex]{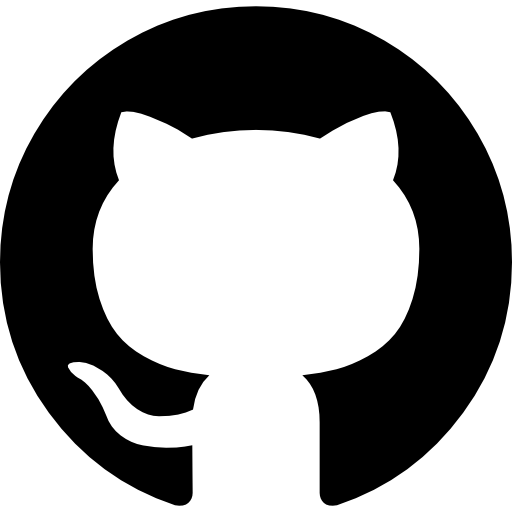}\,\textmd{GitHub Repo}}
\end{tabular}
\normalsize
\end{teaserfigure}

\renewcommand{\shortauthors}{L. Guo, W. Tao, R. Jang, Y. Wang, J. Chen, X. Liu, Y. Ma, M. Mao, H. Zhang, Z. Zheng}

\begin{abstract}
The GitHub issue resolution task aims to resolve issues reported in repositories automatically. With advances in large language models (LLMs), this task has gained increasing attention, and several benchmarks are proposed to evaluate the issue resolution ability of LLMs. However, existing benchmarks have three main limitations. First, current benchmarks focus on a single programming language, limiting the evaluation of issues from repositories across different languages. Second, they usually cover a narrow range of domains, which may fail to represent the diversity of real-world issues. Third, existing benchmarks rely solely on textual information in issue descriptions, overlooking multimodal information such as images in issues. In this paper, we propose \Swebenchx, a \underline{G}itHub \underline{I}ssue \underline{R}eso\underline{L}ution benchmark that is multilingual, multimodal, and multi-domain.  \Swebenchx includes 959 task instances, which are collected from repositories across four programming languages (i.e., Python, JavaScript, TypeScript, and Java) and eight different domains. Our evaluation shows that current LLMs show limited performances on \Swebenchx. Notably, the best-performing model, \GPT, resolves only 8.6\% of the issues. Besides, we find that current LLMs struggle to resolve issues requiring understanding images. The best performance is achieved by \Claude, which resolves only 10.5\% of the issues with image information. Finally, we analyze the reasons behind current LLMs' failure on \Swebenchx, providing insights for future improvements.

\end{abstract}

\ccsdesc[300]{Software and its engineering~Software maintenance tools}

\keywords{Github Issue Resolution, Benchmark, Large Language Models}

\maketitle


\section{Introduction}

The GitHub issue resolution task aims to automatically resolve a wide variety of issues proposed by developers in code repositories. These issues include diverse tasks such as fixing bugs, adding new features, refactoring code, writing documentation, etc~\cite{BissyandeLJRKT13,TaoZWZWZ24}. Effectively addressing these issues is crucial for maintaining and evolving real-world software systems. With the development of large language models (LLMs), 
this task has gained increasing attention~\cite{xia2024agentless,liu2024marscode,tao2024magis,yang2024swe,chen2024coder,zhang2024autocoderover,arora2024masai,wang2024opendevin,ma2024understand}.

Several benchmarks~\cite{zan2024swe,jimenez2023swe,openai2024swe,aleithan2024sweplus} currently exist for the GitHub issue resolution task. SWE-bench~\cite{jimenez2023swe} is the first benchmark in this area, consisting of 2,294 real-world issues from 12 Python repositories. Subsequently, OpenAI proposes SWE-bench Verified~\cite{openai2024swe}, a subset of SWE-bench with 500 instances verified by experienced developers to provide a more robust evaluation. Additionally, SWE-bench-java~\cite{zan2024swe} is introduced to extend the task to the Java programming language, including 93 task instances from 6 Java repositories.  Although various benchmarks have been proposed, there are still some limitations that prevent them from fully capturing the diversity of real-world issue resolution tasks. The primary limitations are as follows:


\begin{itemize}[left=10pt]
    \item \textbf{L1: Focusing on a Single Programming Language.} Existing benchmarks typically focus on issues from repositories in a single programming language, such as Python or Java, which limits their capacity to assess the LLMs' ability to resolve issues across multiple programming languages.


    \item \textbf{L2: Limited Repository Diversity.} Current benchmarks, such as SWE-bench~\cite{jimenez2023swe}, largely rely on issues from a limited range of domains, e.g., scientific computing, machine learning, and visualization. To better represent real-world issue resolution tasks, more repositories from a wider variety of domains are needed.

   \item \textbf{L3: Ignoring Multimodal Information.} Previous benchmarks focus solely on textual information in issue descriptions, overlooking multimodal information. Users often use images, such as screenshots of error messages or debugging outputs, to help illustrate issues more clearly. 
   Without incorporating this information, LLMs may have difficulty fully understanding the issue, leading to inaccurate evaluations. 

\end{itemize}

In this paper, we present \textbf{\Swebenchx}, a benchmark for \textbf{G}itHub \textbf{I}ssue \textbf{R}eso\textbf{L}ution that incorporates multiple aspects of diversity in programming languages, repository domains and modality of input information. 
We first select four most popular programming languages (i.e., Python, TypeScript, JavaScript, and Java) as target languages,\footnote{\url{https://github.blog/news-insights/research/the-state-of-open-source-and-ai/}} and collect a candidate list of the most widely used repositories based on download counts from package management tools (i.e., pip~\cite{pip}, npm~\cite{npm}, maven~\cite{maven}). From this candidate list, we select 15 repositories 
from various domains and collect real-world issues from these repositories to construct task instances. After validating the collected task instances, we obtain a total of 959 task instances covering a variety of domains 
in four programming languages (addressing \textbf{L1} and \textbf{L2}). In addition to textual information in issue descriptions, we find some issues include other modalities. For instance, users use images, such as screenshots of error messages, to describe issues more clearly.  We manually examine issue descriptions on \Swebenchx and identify 19 instances containing images providing crucial information for resolving the issues. 
Moreover, we find that users sometimes share website links to online code platforms, which typically contain code reproducing the issue.  We annotate these links in the dataset, offering future researchers the opportunity to explore how to leverage such resources to enhance issue resolution performance (addressing \textbf{L3}).




We evaluate state-of-the-art LLMs on \Swebenchx, and the results show that the best-performing method, \GPT with the Agentless-X approach,\footnote{Because the Agentless~\cite{xia2024agentless} method is designed for Python language, we extend this method to other programming languages without changing the key design of this method. The multilingual version of this method is called Agentless-X.} resolves only 8.6\% of issues, highlighting the challenges of resolving issues from repositories in multiple languages. Additionally, we observe that the Agentless-X method performs worse on TypeScript and JavaScript tasks than Java and Python. Besides, we evaluate two advanced LLMs with visual abilities in the task instances with images as visual inputs, with \Claude achieving a best resolve rate of only 10.5\% and \GPT achieving just 1.6\%. The results show that both LLMs show limited performance on issues requiring understanding images.




Finally, we analyze the reasons why LLMs fail to resolve issues on \Swebenchx. First, we observe that when using the Agentless method, \Claude often struggles to generate outputs that follow the specified format outlined in the prompt. This formatting issue leads to intermediate results that cannot be parsed correctly, preventing the generation of patches in the next stage. As a result, \Claude achieves a low issue resolve rate of only 1.9\% on \Swebenchx. However, we find that the resolve rate increases to 7.4\% by modifying the prompt simply. Second, we find that the current LLMs have a significantly lower resolve rate on issues that require modifications across multiple files compared to those requiring single-file changes. Besides, for issues needing multi-file modifications, we find models tend to modify a single file, highlighting a limitation in the cross-file issue resolution capabilities of LLMs.

In summary, our contributions are as follows:

\begin{itemize}[left=10pt]
    \item We introduce \Swebenchx, a GitHub issue resolution benchmark with multi-aspect diversity in programming languages, repository domains and modality of input information. 

    \item We evaluate LLMs' issue resolving abilities on \Swebenchx, revealing that current models demonstrate limited overall performance across multiple programming languages.
    
    \item We evaluate LLMs' performance on issues that require visual information, revealing their limited capability in resolving issues with multimodal inputs.

    \item We conduct an analysis to investigate the reasons why LLMs fail to resolve issues, providing insights for improving issue resolving performance of LLMs in the future.
\end{itemize}


\begin{table}[t]
\centering
\footnotesize
\setlength\tabcolsep{3pt}
\caption{Overview of existing GitHub issue resolution datasets. }
\tabmargin
    \begin{tabular}{lcccccccc}
    \toprule
    \multirow{2}{*}{\textbf{Dataset}} & \multirow{2}{*}{\textbf{Year}} & \multicolumn{4}{c}{\textbf{Programming Languages}} & \multirow{2}{*}{\textbf{\# Repositories}} & \multirow{2}{*}{\textbf{\# Instances }} & \textbf{w/ Visual} \\
    \cmidrule(r){3-6}
    && \bf Python & \bf Java & \bf TypeScript & \bf JavaScript & & & \bf Input\\
    \midrule
    SWE-bench~\cite{puri2021codenet} & 2023 & $\checkmark$ & $\times$ & $\times$& $\times$& 12 & 2,294 & $\times$ \\
    SWE-bench-Java~\cite{ahmad2021avatar} & 2024 & $\times$& $\checkmark$ & $\times$& $\times$&6 & 91 & $\times$  \\
    SWE-bench-Verified~\cite{liu2024your} & 2024 & $\checkmark$ &$\times$ &$\times$ &$\times$ &12& 500 & $\times$  \\
    \midrule

    \Swebenchx &2024& $\checkmark$ & $\checkmark$ & $\checkmark$ &$\checkmark$ & 15 & 959 & $\checkmark$  \\
    \toprule
\end{tabular}
\label{table:datasets}
\end{table}%

\secmargin
\section{Background and Related Work}

\begin{figure}[t]
    \centering
    \includegraphics[width=0.85\linewidth]{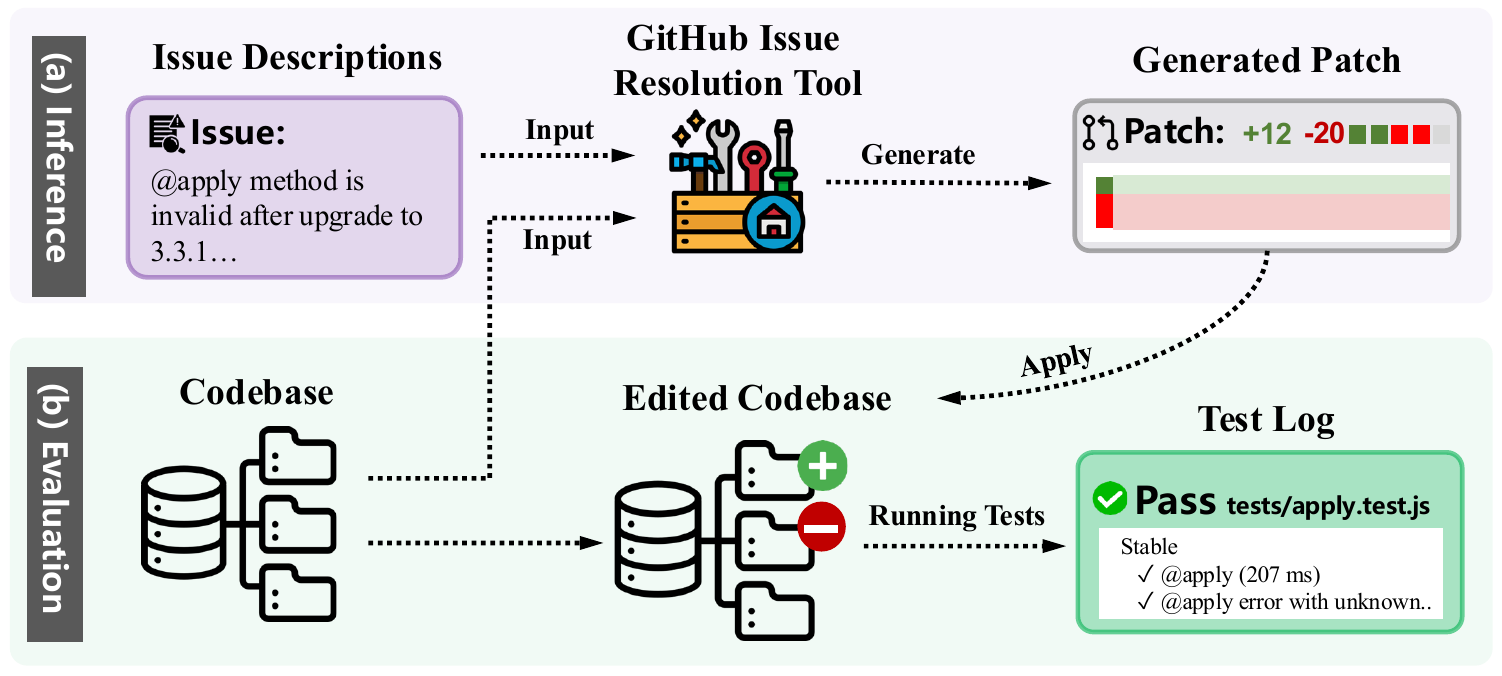}
    \figmargin
   \caption{Overview of the evaluation pipeline of GitHub issue resolution.}
    \label{fig:TaskDefinition}
\end{figure}

\subsection{GitHub Issue Resolution}
The GitHub issue resolution task aims to resolve issues reported in the GitHub repository automatically~\cite{jimenez2023swe,zan2024swe}. 
When evaluating the issue resolution ability of tools, we follow the pipeline as shown in Figure~\ref{fig:TaskDefinition}. During the inference stage, the tool receives the issue descriptions along with the corresponding codebase and generates a patch that contains all necessary code changes to the original codebase. In the evaluation stage, the generated patch is applied to the codebase, and all test cases related to this issue are run. After obtaining the test log, we check the statuses of the test cases. The issue is considered resolved if all related test cases pass successfully. Some key concepts are listed below:

\begin{itemize}
    \item \textbf{Issue Descriptions:} descriptions about reported issues in the format of texts or images. 
    \item \textbf{Generated Patch:} a generated file including all code changes in the code repository to resolve reported issues. All code changes are formatted in the GitHub diff format.
    \item \textbf{Codebase:} the code repository containing the reported issue and all relevant code files necessary for resolving the issue.
    \item \textbf{Test Log:} a log file containing all test results, which is used to verify the correctness of results.
\end{itemize}


\subsection{LLM-based Methods for GitHub Issue Resolution}
\label{sec:methodBackground}




With the development of LLMs, many researchers attempt to explore their potential for automating GitHub issue resolution. Jimenez et al.~\cite{jimenez2023swe} are the first to use LLMs, such as GPT-4, to resolve issues in SWE-bench~\cite{jimenez2023swe}. However, the performances of LLMs were limited at that time, with the best-performing model, Claude-2, resolving only 1.96\% of issues. Inspired by the success of agent frameworks in software engineering tasks~\cite{huang2024agents,liu2024large,feldt2023towards}, some researchers propose agent-based methods~\cite{tao2024magis,zhang2024autocoderover,yang2024swe,chen2024coder,liu2024marscode,wang2024opendevin,bouzenia2024repairagent,ruan2024specrover,zhang2024diversity} to enhance the performance of LLMs in issue resolution tasks. Tao et al.~\cite{tao2024magis} propose the first multi-agent-based issue resolution framework, MAGIS, to improve the issue resolution ability of LLMs. Yang et al.~\cite{yang2024swe} propose the SWE-agent framework to build an LLM-based agent that can autonomously utilize designed tools to resolve issues. Besides, some works also utilize existing software engineering techniques~\cite{wong2016survey,jones2005empirical,lou2020can} to improve the performance of LLMs. Zhang et al. propose the AutoCodeRover~\cite{zhang2024autocoderover} method, which enhances the localization of edited code by analyzing the structure of the repository.  Inspired
by existing LLM-based APR tools~\cite{xia2023automated,xia2023keep,bouzenia2024repairagent,hidvegi2024cigar,chen2024flakiness,zhang2024acfix}, Xia et al.~\cite{xia2024agentless} propose the Agentless method, which uses a hierarchical process to find the edit location. Compared with agent-based frameworks, software engineering oriented methods can achieve comparable performance and cost less.

Following previous studies~\cite{anthropic2024building,ouyang2024repograph,pan2024training}, we can divide current LLM-based issue resolution methods into three types: RAG-based method, LLM workflow-based method, and LLM agent-based method. The RAG-based methods use a retriever to directly retrieve similar code files from the codebase to enhance issue resolution. However, these methods, such as the BM25 retrieval-based method, demonstrate limited performance on this task~\cite{jimenez2023swe}. The second type, the LLM workflow-based method, breaks down the issue resolution process into predefined stages, where LLMs follow a fixed sequence to complete tasks. For example, Agentless~\cite{xia2024agentless} employs a structured workflow with a localization stage to retrieve relevant code and a patch generation stage to generate a final patch. Finally, the LLM agent-based method, such as AutoCodeRover~\cite{zhang2024autocoderover} and SWE-agent~\cite{yang2024swe}, allow LLMs to autonomously interact with the codebase, using tools to collect key information and resolve tasks dynamically. 

\subsection{Benchmarks for GitHub Issue Resolution}
Recently, several benchmarks have been introduced to evaluate the issue resolution abilities of LLMs, as summarized in Table~\ref{table:datasets}. Jimenez et al.~\cite{jimenez2023swe} propose SWE-bench, the first GitHub issue resolution benchmark, comprising 2,294 resolved issues from 12 Python repositories. Subsequently, OpenAI releases SWE-bench Verified~\cite{openai2024swe}, a subset of SWE-bench containing 500 task instances verified by experienced developers to ensure correctness. Zan et al.~\cite{zan2024swe} further introduce SWE-bench-java, with 91 task instances from 6 Java repositories. 

However, existing benchmarks have some limitations. Unlike other software engineering tasks~\cite{zheng2023towards}, where benchmarks evaluate LLMs' coding abilities across multiple languages~\cite{zheng2023codegeex,microsoft2023multipl,TaoWSDH0ZZ21, athiwaratkun2023multilingual,zhang2023humanevalx,TaoWSDHZZZ22,yu2024codereval,yan2023codetransocean,yan2023codescope,zheng2024towards,zheng2024well,chen2024rmcbench}, current issue resolution benchmarks typically focus on a single language, which restricts evaluation diversity.  Additionally, benchmarks like SWE-bench~\cite{jimenez2023swe} collect data from limited domains, such as scientific computing, machine learning, and visualization. Moreover, current benchmarks focus solely on text information in issue descriptions, overlooking multimodal data such as images. To address these limitations, we propose \textbf{\Swebenchx}, a \textbf{G}itHub \textbf{I}ssue \textbf{R}eso\textbf{L}ution benchmark with \textbf{Omni}-aspect diversity in programming languages, repository domains and modality of input information.



\secmargin
\section{\Swebenchx Construction}

\begin{figure}
    \centering
    \includegraphics[width=0.99\linewidth]{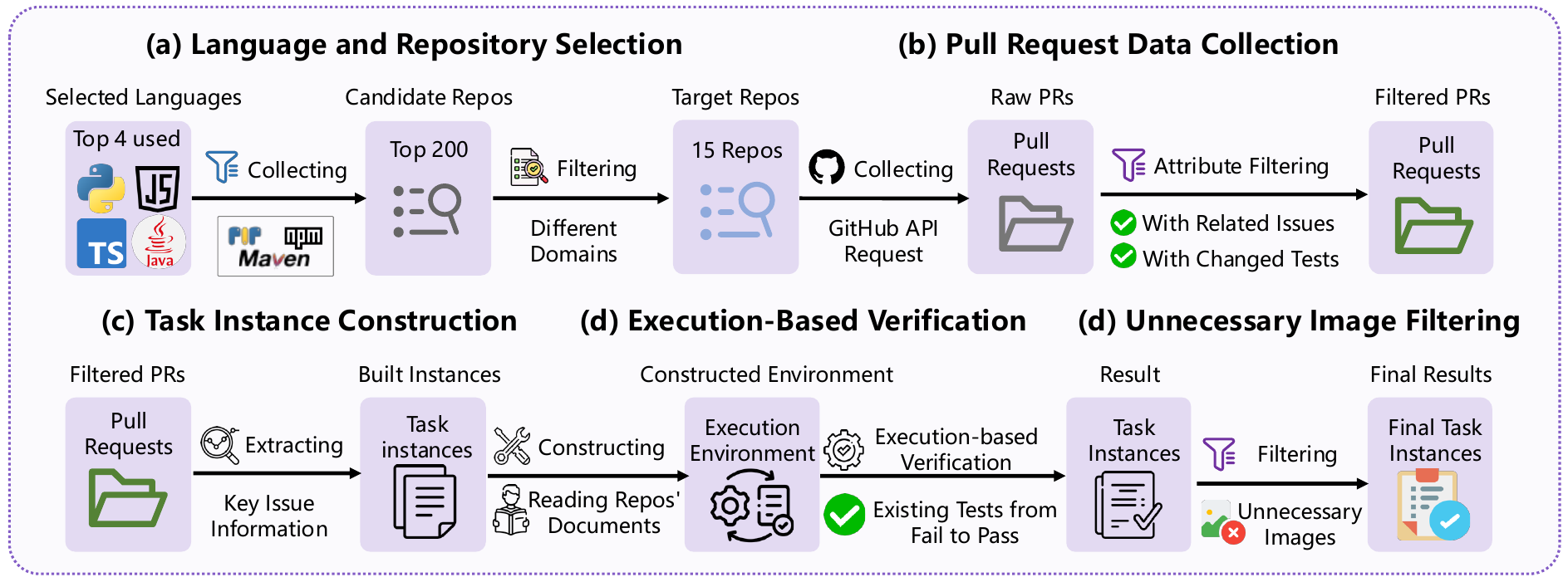}
    \figmargin
    \caption{Overview of benchmark construction.}
    \label{fig:DataCollection}
\end{figure}

In this section, we introduce the process of building  \Swebenchx. As shown in Figure~\ref{fig:DataCollection},  this process includes five stages: (a) language and repository selection, (b) pull request data collection, (c) task instance construction, (d) execution-based verification and (e) unnecessary image filtering.

\subsection{Language and Repository Selection}

\subsubsection{Language Selection.} Before building our multilingual benchmark, we first select widely used programming languages as target languages. Based on the latest GitHub report,\footnote{\url{https://github.blog/news-insights/research/the-state-of-open-source-and-ai/}} JavaScript, Python, TypeScript, and Java are among the most popular languages, reflecting their extensive use in the developer community. Besides, these languages are widely used in different domains: Python is common in AI and data science, JavaScript and TypeScript are crucial for web development, and Java is widely applied in large-scale systems and backend development. Considering the popularity and significance of these languages, we choose them as our target languages.

\subsubsection{Repository Selection.} 

To ensure the popularity of the target repositories, we use the download counts from language-specific package management tools. For Python, we select repositories based on pip download counts. For Java, we use Maven, and for JavaScript and TypeScript, we rely on npm. For each language, we select the top 200 repositories with the highest download counts. Next, using the GitHub API~\cite{github-rest-api}, we collect the tags from these repositories. Based on the tags, we select 15 popular repositories from different domains as our target repositories. This approach ensures that our selection covers diverse fields and focuses on highly used repositories.

\subsection{Pull Request Data Collection} 
\subsubsection{Collection of Pull Requests.} In GitHub repositories, developers submit pull requests to report potential issues and resolve existing issues. In this stage, we use the GitHub Developer API~\cite{github-rest-api} to collect pull requests from each repository. We then retain only those pull requests in the merged state. This is because merged pull requests have been generally reviewed by the repository maintainers and successfully integrated into the codebase. Additionally, we set a cutoff date of July 31, 2024, and only keep pull requests that were merged before this date, using this as a starting point for future data collection.




\subsubsection{Attribute-Based Filtering.}
To make that each collected data contains issue descriptions and tests to verify the correctness of submitted solutions, following the approach of SWE-bench~\cite{jimenez2023swe}, we use the attribute-based filtering method to filter the collected pull requests data further:

\begin{itemize}[left=10pt]
    \item \textbf{Keep PRs resolving at least one issue.} In the GitHub repository, when submitting a pull request that resolves some issues, the developer uses statements like ``fix \#403'' in the title or body to indicate which issues are resolved. Following the implementation code of SWE-bench,\footnote{\url{https://github.com/princeton-nlp/SWE-bench/blob/main/swebench/collect/utils.py}} we first gather all text content from the title, body, and commit messages of each pull request. We then extract each issue number (e.g., ``\#403'') from this aggregated content using regular expressions. Finally, we filter out pull request data without any relevant issue number.
    \item \textbf{Keep PRs with test files changed.}
When fixing a bug or adding a new feature, developers often submit a pull request containing code changes in test files. These test files offer a great solution to verify whether the bug is fixed or the feature is implemented successfully. Following SWE-bench,$^4$ we first identify test files in the changed files of a pull request by checking if their full paths contain keywords like ``test'' or ``testing''. Then, we filter out pull request data without any test file changed.
\end{itemize}

\begin{figure}
    \centering
    \includegraphics[width=0.95\linewidth]{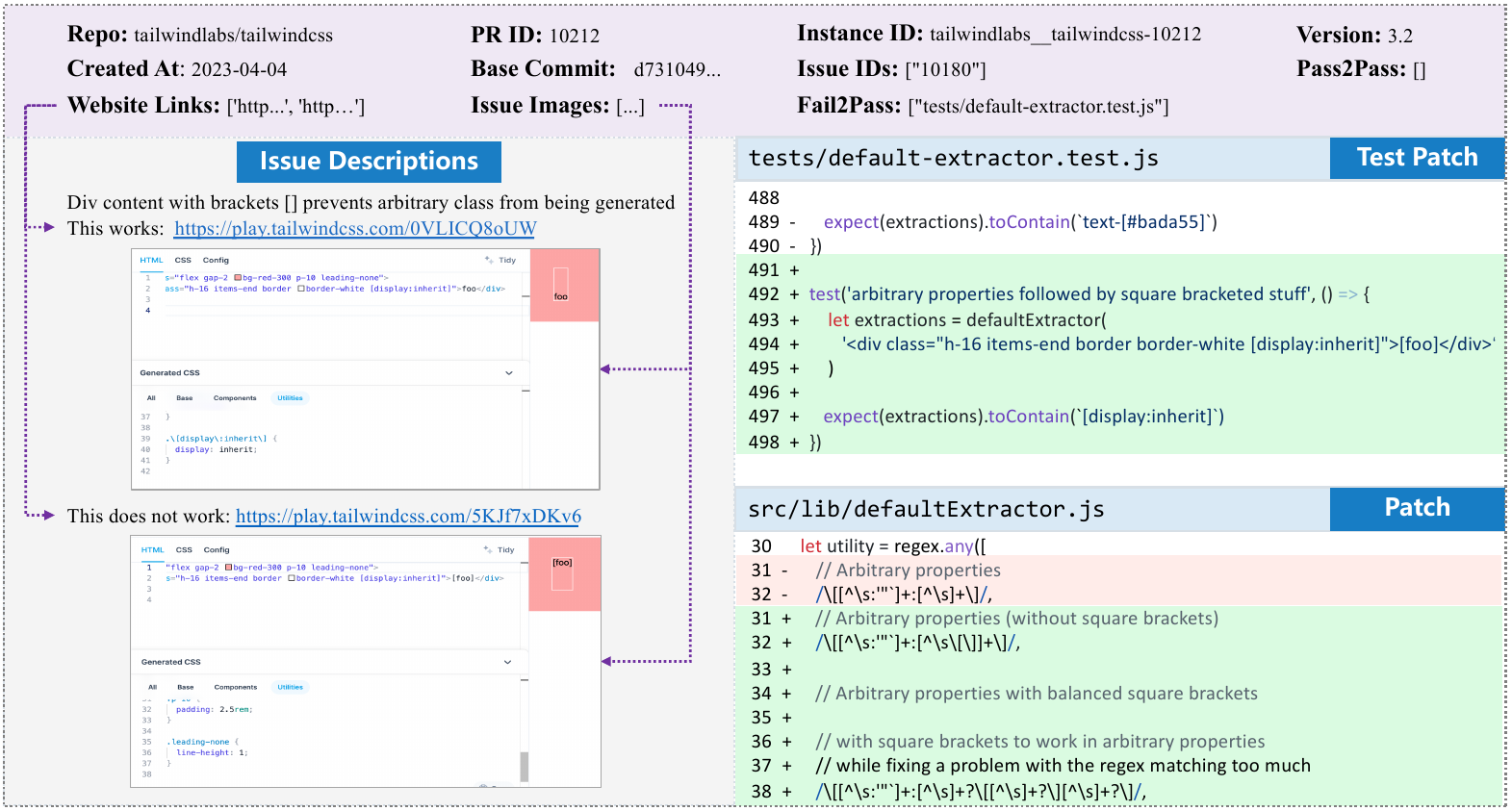}
    \figmargin
    \caption{An example of task instance \texttt{tailwindlabs__tailwindcss-10212}.} 
    %
    \label{fig:TaskInstanceExample}
\end{figure}

\subsection{Task Instance Construction}
After collecting raw data of pull requests, we build task instances using GitHub Developer API~\cite{github-rest-api} to obtain key information from each pull request. An example of a task instance is shown in Figure~\ref{fig:TaskInstanceExample}. Here, we introduce the attributes of this task instance:

\begin{itemize}[left=10pt]
    \item \textbf{Repo.} This attribute refers to which repository the task instance belongs to. In Figure~\ref{fig:TaskInstanceExample}, this instance is from the repository ``tailwindlabs/tailwindcss''. 
    \item \textbf{PR ID.} This attribute refers to which pull request the task instance belongs to. Each pull request has a unique ID, such as 10212.
    \item  \textbf{Instance ID.} This attribute is a unique identifier for the task instance, composed of ``Repo'' and ``PR ID''. For example, the ``tailwindlabs\_\_tailwindcss-10943'' means this task instance is obtained from pull \#10943 in the repository ``tailwindlabs/tailwindcss''.
    \item  \textbf{Created At.} This attribute refers to the date when the pull request was created.
    \item \textbf{Issue IDs.} This attribute refers to the ID numbers of the issues that are resolved by the pull request. For example, the issue number of instance ``tailwindlabs\_\_tailwindcss-10943'' is 10937, which means this pull request resolves the issue \#10937.
    \item \textbf{Issue Descriptions.} This attribute refers to the text descriptions of issues related to the task instance. For example, as shown in Figure~\ref{fig:TaskInstanceExample}, the problem statement describes unexpected results of running programs. In the evaluation, this problem statement serves as the input of the task instance. Following the approach of  SWE-bench, we extract and concatenate the issue's title and body to form the content of the problem statement.
    \item \textbf{Issue Images.} This attribute refers to images in issue descriptions.  As shown in Figure~\ref{fig:TaskInstanceExample}, this user attaches two images to describe the unexpected results of running programs. In the evaluation, these images can provide visual information for resolving issues. Since GitHub uses URLs to display images, we extract URLs containing ``png'' or ``jpg'' from the issue body of task instances. We then manually check whether these images are relevant to the content of the issue. Because not all task instances include images, this is an optional attribute.
    \item \textbf{Website Links.} This attribute refers to some website links users share to help describe reported issues. As shown in Figure~\ref{fig:TaskInstanceExample}, this user shares two links to an online code execution platform, where the reproduced code and execution results are presented. These links provide execution environments for developers to understand issues by debugging.  We manually check website links in each issue and keep website links that are crucial for resolving this issue. This is an optional attribute because not all task instances include such links. 

    
    \item \textbf{Version.} This attribute refers to the version of the repository when the pull request was created. Since different versions of the code repository may require different environment setups, this information helps construct the appropriate code environment for each task instance. Considering that version information is typically updated in configuration files, we locate the configuration file corresponding to each task instance and extract the version information. For example, Python versions are usually stored in the \texttt{setup.py} file, JavaScript and TypeScript versions in \texttt{package.json}, and Java versions in \texttt{pom.xml}.
    \item \textbf{Base Commit.} This attribute is a unique commit ID that the original pull request is based on. Using the base\_commit, we can revert the GitHub repository to the state before the pull request was applied. This helps us construct the correct code repository environment for the target issues. We extract this information using the GitHub Developer API~\cite{github-rest-api}.
      \item \textbf{Test Patch.} This attribute refers to the code changes in the pull request that are used to test the modified source code. In the evaluation, this content is used to run tests to verify the correctness of the submitted solution. Following the approach of SWE-bench~\cite{jimenez2023swe}, we check the file paths of each hunk in the code changes and retain only those where the file path contains keywords like ``test'' or ``testing''.  Additionally, for Java, we consider any Java file where the file name starts or ends with ``Test'' (e.g., \texttt{TestClass.java} or \texttt{MyClassTest.java)} as part of the test patch.
    \item \textbf{Patch.} This attribute refers to the code changes in the pull request that specifically address resolving the issue. This content provides the ground truth solution for the issue. To collect this data, we first examine each hunk in the code changes and filter out hunks related to testing. Then, we use the Python library Pygments~\cite{pygments} to check the file paths in each hunk and determine if they belong to the source files of the target programming language. For example, in Python repositories, files ending in ``.py'' are recognized as Python source files. This helps us exclude irrelevant files, such as documentation files ending in ``.md''. Additionally, we collect certain configuration files that are crucial for the patch’s correctness. For instance, for JavaScript, we gather changes in \texttt{package.json}, and for Java, we include changes in \texttt{pom.xml}. These specific changes are important for ensuring the patch's correctness.
    
    \item \textbf{FAIL2PASS.} This attribute refers to tests or test cases that are changed from ``fail'' status to ``pass'' status after applying the gold patch. These tests or test cases are used to verify whether the submitted solution can resolve issues in the task instance.
    \item \textbf{PASS2PASS.} This attribute refers to tests or test cases that maintain a ``pass'' status both before and after applying the gold patch. This attribute can be used to verify whether the submitted solution does not change the function of the original source code. 
\end{itemize}


\subsection{Execution-Based Verification} 
In this section,  we conduct execution-based verification to make each task instance have paired tests to check the correctness of submitted solutions. First, we construct an execution environment for each task instance and verify its correctness. Second, following the approach of SWE-bench~\cite{jimenez2023swe}, we conduct execution-based filtering to filter task instances without FAIL2PASS tests or test cases. The remaining task instances are considered valid data on \Swebenchx.


\subsubsection{Environment Construction.} 
Before conducting execution-based filtering, we create an isolated execution environment for each task instance using Docker~\cite{docker} to minimize potential conflicts across environments. For each task instance, we define all setup commands in a \texttt{Dockerfile} and bash scripts, which are then executed to set up the Docker environment for each instance.

The execution environment includes the codebase before the issue was resolved and its required dependencies. For each task instance, we first use \texttt{Git}~\cite{git} to clone the repository. Then, with the task instance's ``Base Commit'' attribute, we use \texttt{Git} to revert the repository to the state before the pull request addressing the issue was submitted. To set up the correct dependencies, we refer to development documentation in the codebase, such as \texttt{README.md} and \texttt{CONTRIBUTING.md}, which often provide guidance on how to build dependencies. Following these instructions, we install the necessary dependencies for the codebase. The process described above is written into the \texttt{Dockerfile} and bash scripts to automate the construction of the docker environment.




\subsubsection{Environment Verification.}After building the execution environment successfully, we need to verify its correctness. In this step, our aim is to ensure that all tests in the test patch can be passed after applying the gold patch to the environment. Firstly, we update test files by applying the test patch to the codebase. Then, we apply the gold patch to the codebase and run relevant tests to obtain test results. We consider the constructed environment successfully verified only if all test cases pass. Otherwise, we think this environment needs further improvements.
\subsubsection{Execution-Based Filtering}
After building the execution environment, the next step is to verify if the tests can effectively evaluate the correctness of the submitted patch. Ideally, after applying the solution (gold patch), we expect the status of certain test cases to change from ``fail'' to ``pass'', indicating that the solution has addressed specific issues. We refer to these as FAIL2PASS test cases, which serve to verify that the patch has resolved the intended issue in the task instance. Test cases that maintain a 'pass' status throughout this process are called PASS2PASS test cases. These help ensure that the submitted patch does not change the original functionality of the code.

Following the approach of SWE-bench~\cite{jimenez2023swe}, we first run the relevant tests before and after applying the gold patch to observe whether there exist FAIL2PASS test cases. Then, We only retain task instances that contain at least one FAIL2PASS test case. Finally, we manually check whether the FAIL2PASS test cases match the tests  added or modified in the test patch, ensuring higher reliability in our evaluation process.

\subsection{Unnecessary Image Filtering}  
After building task instances, we manually check images in each task instance to determine whether these images are essential for resolving issues. For example, while some users utilize screenshots of error messages to describe issues, these error messages may already be included in the text of the issue descriptions. In such cases, we filter these unnecessary images, as they do not provide crucial information for resolving issues.
The results of the filtering process are presented in Table~\ref{table:filteringProcess}. Before filtering, the dataset contains 39 task instances with an average of 1.7 images per instance, covering ten repositories across all four languages. After filtering out unnecessary images, \Swebenchx remains 19 instances with an average of 1.8 images per instance, now covering seven repositories across three languages (e.g., Python, TypeScript, and JavaScript).



\begin{table}[t]
  \centering
  \footnotesize
  \setlength\tabcolsep{1.5pt}
  \caption{Results of filtering unnecessary images.}
  \label{table:filteringProcess}
  \tabmargin
    \begin{tabular}{ccccccc}
    \toprule
    \multirow{2}{*}{\textbf{Language}} & \multicolumn{3}{c}{\textbf{Before Filtering}} & \multicolumn{3}{c}{\textbf{After Filtering}} \\
    \cmidrule(r){2-4} \cmidrule(r){5-7}
    & \textbf{\# Instances} & \textbf{\# Avg Images} & \textbf{\# Repos} & \textbf{\# Instances} & \textbf{\# Avg Images} & \textbf{\# Repos} \\
    \midrule
    
    \textbf{All} & 39  & 1.7 & 10 &  19 & 1.8 & 7 \\
    \midrule
    \quad Python & 7 & 1.3 & 2 & 2 & 1.0 & 2 \\
    \quad TypeScript  & 15 & 1.7 & 2 & 12 & 1.9 & 2 \\
    \quad JavaScript  & 15 & 1.8 & 3 & 5 & 1.8 & 3 \\
    \quad Java  & 2 & 1.5 & 2 & 0 & 0.0 & 0 \\

    \bottomrule
    \end{tabular}

\end{table}
\secmargin
\section{\Swebenchx}
\subsection{Statistics of \Swebenchx}
After constructing the benchmark, \Swebenchx includes 959 instances collected from 15 repositories, covering four programming languages: Python, JavaScript, TypeScript, and Java. The details of the repositories included in the dataset are presented in Table~\ref{table:statitics}, which lists the programming language used, the number of instances, and the license for each repository. The licenses ensure that our dataset can be freely accessed and used for research purposes.

\begin{table}[t]

\centering
\caption{Statistics of \Swebenchx repositories.}
\footnotesize
\tabmargin
\begin{tabular}{lccc}
\toprule
\textbf{Repository} & \textbf{\# Instances} & \textbf{\# Languages} & \textbf{License} \\
\midrule
python/mypy & 189 & Python & BSD 3-Clause \\
pyca/cryptography & 21 & Python & BSD 3-Clause  \\
dateutil/dateutil & 36 & Python & BSD 3-Clause \\
tqdm/tqdm & 23 & Python & MIT \\
statsmodels/statsmodels & 76 & Python & BSD 3-Clause \\
redis/redis-py & 29 & Python & MIT \\
iamkun/dayjs & 93 & JavaScript & MIT \\
prettier/prettier & 119 & JavaScript & MIT \\
webpack/webpack & 58 & JavaScript & MIT \\
jestjs/jest & 31 & TypeScript & MIT \\
babel/babel & 79 & TypeScript & MIT \\
tailwindlabs/tailwindcss & 100 & TypeScript & MIT \\
netty/netty & 54 & Java & Apache-2.0 \\
google/gson & 21 & Java & Apache-2.0 \\
assertj/assertj & 30 & Java & Apache-2.0 \\
\toprule
\end{tabular}

\label{table:statitics}
\end{table}

We also analyze the characteristics of task instances, as summarized in Table~\ref{table:dataAnalysis}. We can find that task instances contain a long issue text, including 194 words on average. The codebase includes 995 files and 257K lines of code on average, showing the complexity of the codebase. Besides, the gold patches require modifications across 46 lines, 2.2 functions, and 1.2 files on average. These characteristics highlight the complexity of GitHub issue resolution tasks.

\begin{table}[t]
\caption{Data analysis for our benchmark.} 
\label{table:dataAnalysis}
\tabmargin
\centering
\footnotesize
\begin{tabular}{cccc}
\toprule
& &Mean&Max \\
\midrule
Issue Text&Length (Words)&194&1,685 \\
\midrule
\multirow{2}{*}{Codebase}&\# Files (non-test)&995&3,016 \\
&\# Lines (non-test)&257K&1,010K \\
\midrule
\multirow{3}{*}{Gold Patch}&\# Lines edited& 46.3 & 1,657\\
&\# Files edited&1.2&30 \\
&\# Func. edited&2.2&131 \\
\midrule
\multirow{2}{*}{Tests}&\# Fail to Pass&3.3&1,056 \\
&\# Total&135.1&4,820 \\
\bottomrule
\end{tabular}
\end{table}


\subsection{Diversity of \Swebenchx}  
In this section, we introduce the diversity of \Swebenchx from three aspects: diversity of programming languages, diversity of repository domains, and diverse modality of input information.

\subsubsection{Diversity of Programming Languages}

Our dataset contains issues from repositories in four widely used programming languages: Python, Java, JavaScript, and TypeScript. Python is known for its readability and extensive library support and is widely used in data science and machine learning. Java is commonly used in large-scale applications due to its great performance and robustness. JavaScript dominates web development with its dynamic capabilities. TypeScript is a statically typed superset of JavaScript, which adds type safety to JavaScript's flexibility.
Resolving issues from repositories across different programming languages requires distinct background knowledge. For each issue, we use its gold patch as a reference to analyze the required background knowledge. Each label of background knowledge can be found in developer documentation of these four languages~\cite{jsDocs,tsDocs,javaDocs,pythonDocs}. The analysis results are shown in Table~\ref{table:languageFeatures}. For example, we find issues in the JavaScript repository require background knowledge of both JavaScript and TypeScript (e.g., ``Arrow Functions'' and ``interface''). By identifying this rich background knowledge, our dataset enables researchers to design issue resolution methods applicable across different programming languages.

\begin{table}[t]
\scriptsize
\centering
\caption{Categories of background knowledge in different programming languages contained in gold patch.}
\label{table:languageFeatures}
\tabmargin
\begin{tabular}{c l}
\toprule
    \bf{Language of Repository}  & \bf{Background Knowledge Contained in Gold Patch} \\
\midrule
    Python  & ``Decorator''; ``Generator''; ``Iterator''; ``Reflection''; ``Wraps''\\
    JavaScript & ``Anonymous Functions''; ``Arrow Functions''; ``Event-driven''; ``Interface''; \\
    TypeScript  & ``Anonymous Functions''; ``Arrow Functions''; ``Enum''; ``Generics''; ``Interface''; ``Type Aliases'';  \\
    Java  & ``Abstract Classes''; ``Annotations''; ``Generics''; ``Inheritance''; ``Interface'';  ``Reflection'';  \\
\bottomrule
\end{tabular}

\end{table}

\subsubsection{Diversity of Repository Domains.} In this section, we present the diverse range of domains covered by the repositories included in our benchmark. We first collect a brief introduction from the GitHub summary of each repository and label them with relevant domain tags. As shown in Table~\ref{table:repoSummary}, our benchmark covers domains as follow:

\begin{itemize}[left=10pt]

\item \textbf{Code quality (35.2\%).} Code quality tool is widely used to help developers improve code readability and maintainability. For instance, ``Prettier'' automatically formats code to ensure consistent style, making it more readable and maintainable.

\item \textbf{Web development (24.7\%).}  
Web development tools like ``tailwindcss'' are widely used for building modern, responsive user interfaces, facilitating efficient web development and ensuring compatibility with modern browsers and frameworks.

\item \textbf{Time tools (13.5\%).}  
Repositories such as ``dateutil'' provide crucial utilities for handling time and date operations. These libraries are widely used in various applications, particularly in scheduling and time-sensitive tasks across industries.

\item \textbf{Network tools (8.7\%).}  
Repositories like ``redis-py'' provide fundamental tools for managing databases and building scalable network applications. These libraries are key to developing high-performance, distributed systems.

\item \textbf{Statistical modeling (7.9\%).}  The ``statsmodels'' library provides rich statistical and econometric tools essential for data-driven decision-making and financial trend modeling. Maintaining this repository requires rich knowledge of statistics and econometrics.


\item \textbf{Developer utility (4.4\%).}  
Developer Utility provides tools to enhance development productivity. For instance, ``gson'' simplifies JSON parsing and serialization in Java, while ``tqdm'' offers progress bars for Python, helping developers efficiently monitor and manage tasks.

\item \textbf{Test tool (3.2\%).}  
Test tools such as ``jest'' automate software testing, ensuring code correctness, reliability, and performance by detecting bugs and verifying expected functionality.

\item \textbf{Cryptography (2.4\%).}  
The ``cryptography'' repository offers essential cryptographic primitives and recipes, playing a critical role in securing communications and data, especially in sensitive fields like cybersecurity and financial services.

\end{itemize}

\begin{table}[t]
\centering
\footnotesize
\caption{Repository summary and domain tags.}
\tabmargin
\resizebox{0.85\textwidth}{!}{%
\begin{tabular}{lll}
\toprule
\textbf{Repository} & \textbf{Summary} & \textbf{Domain Tag} \\
\midrule
python/mypy & Optional static typing for Python  & Code quality \\
prettier/prettier & Opinionated code formatter for JavaScript & Code quality \\
assertj/assertj & Library providing typed assertions for Java & Code quality \\
webpack/webpack & A bundler for javascript to pack modules & Web development\\
babel/babel & Compiler for writing next generation JavaScript. & Web development\\
tailwindlabs/tailwindcss & Utility-first CSS framework for UI development & Web development\\
dateutil/dateutil & Extensions to the standard Python datetime features & Time tool\\
iamkun/dayjs & Lightweight immutable date-time library & Time tool\\
redis/redis-py & Redis python client & Network tool\\
netty/netty & Asynchronous network application framework & Network tool\\
statsmodels/statsmodels & Statistical modeling and econometrics in Python & Statitics modeling\\
tqdm/tqdm & Fast, Extensible Progress Bar for Python and CLI & Developer utility\\
google/gson & Java library for JSON serialization and deserialization &Developer utility\\
jestjs/jest & Testing Framework for JavaScript & Test tool\\
pyca/cryptography & Cryptographic recipes and primitives for Python & Cryptography\\

\toprule
\end{tabular}%
}

\label{table:repoSummary}
\end{table}


\subsubsection{Diversity of Input Information Modalities} Previous benchmarks only focus on text information in issue descriptions, overlooking diverse multimodal information on issues. By manually reviewing issues in our benchmark, we identify three modalities of input: text information, image information, and website information.

\textbf{Textual information.} Text is the most common modality in GitHub issues, where users typically describe issues with details such as environment setup, reproduction code, error messages, etc.

\begin{figure}[t]
    \centering
    \includegraphics[width=\linewidth]{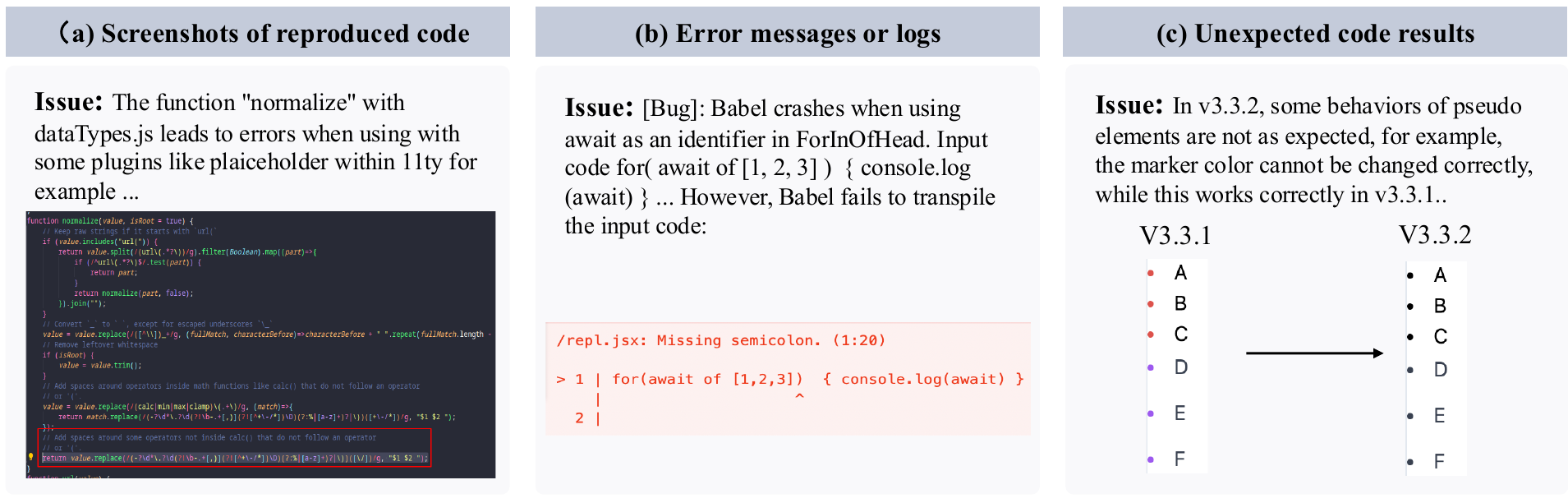}
    \figmargin
    \caption{Examples of images contained in issue descriptions.}
    \label{fig:ImageExample}
\end{figure}

\textbf{Image information.}
In addition to textual information, users often upload images to provide key details of reported issues. After analyzing images contained on \Swebenchx, we classify them into three types based on their purpose, as shown in Figure~\ref{fig:ImageExample}: 

\begin{itemize}[left=10pt]
    \item \textbf{Screenshots of reproduced code.}  Some users prefer to attach screenshots to demonstrate the code that reproduces an issue, as shown in Figure~\ref{fig:ImageExample} (a). 

    \item \textbf{Error messages or logs.}  Some users prefer taking screenshots of error messages or logs instead of copying the text, especially when the error is long or has complex formatting. In Figure~\ref{fig:ImageExample} (b), the user uploads a screenshot to show the error messages.
    

    
    \item \textbf{Unexpected output or behavior of code.} These images are used to report unexpected results or behaviors after running the program.
    Unlike the previous two categories, these images often cannot be easily converted to text and require LLMs to understand the semantics of images with visual ability. For instance, in Figure~\ref{fig:ImageExample} (c), after updating the code version, the user noticed that the colors of the elements were incorrect. They used images to show the unexpected changes in the element colors. 

\end{itemize}
Image information presents new challenges for current LLMs, requiring models with visual abilities to understand images, extract key issue details, and ultimately assist in resolving issues.

\begin{figure}[t]
    \centering
    \includegraphics[width=\linewidth]{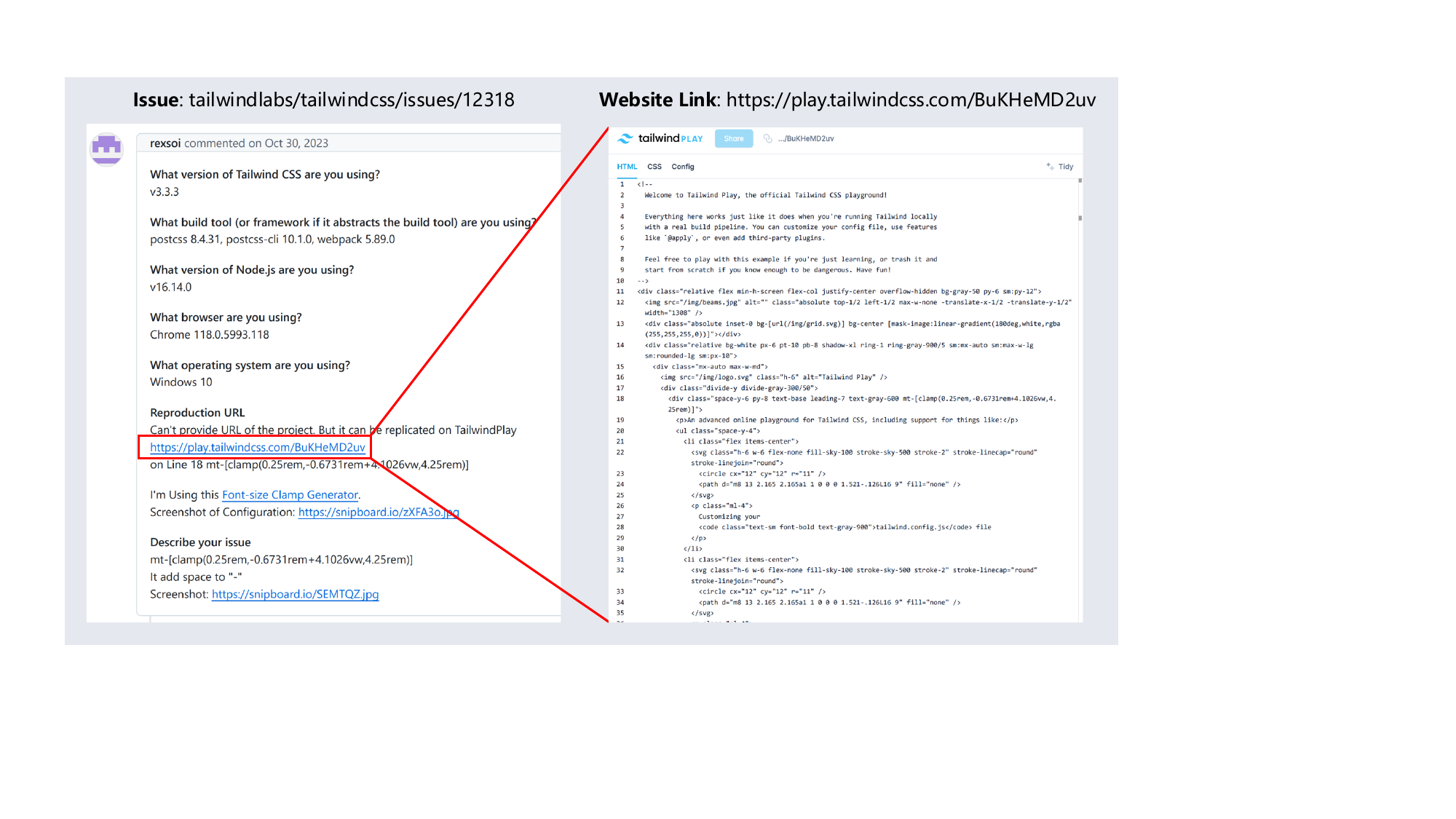}
    \figmargin
    \caption{An example of website links within an issue.}
    \label{fig:url}
\end{figure}

\textbf{Website information.} Website information refers to some website links that provide important details related to an issue.  For example, in the JavaScript and TypeScript GitHub communities, users often share links to online code execution platforms to help others reproduce reported issues. As shown in Figure~\ref{fig:url}, instead of providing the reproduced code directly, a user shares a link in the issue description that directs to ``Tailwind Play'', a website for debugging Tailwind CSS code online. These links allow developers to execute the reproduced code and observe errors in real-time. Unlike textual or image information, website information requires models to have web-browsing capabilities to interact with the website and obtain crucial information about issues.

%

\footnotetext{\url{https://github.com/tailwindlabs/tailwindcss/issues/12318}}

\textbf{Distribution of different modality.} 
We analyze the modalities present in each issue to calculate their distribution. Our analysis shows that all issues include text information, with 24.0\% (230/959) also including website information, 4.1\% (39/959) containing image information, This distribution highlights the diversity of multimodal data on \Swebenchx.
\secmargin
\section{Evaluation}

In this section, we first evaluate the issue resolution ability of three advanced LLMs on \Swebenchx. We then evaluate the performances of LLMs on issues that requires image understanding. Finally, we investigate why LLMs fail to resolve issues on \Swebenchx. 
Note that we only evaluate LLM-based methods here because, according to the SWE-bench leaderboard~\cite{swebench2024leaderboard}, the most advanced tools are primarily LLM-based. However, our benchmark is designed to evaluate the issue resolution ability of any tools, including those that do not rely on LLMs. In summary,  we answer the following research questions: 




\begin{itemize}

    \item \textbf{RQ1:} How do the most advanced LLMs perform on \Swebenchx? 
    \item \textbf{RQ2:} How do the most advanced LLMs perform in resolving issues with images?
    \item \textbf{RQ3:} What are the main reasons for LLM failures on \Swebenchx?
\end{itemize}

\subsection{Experimental Setup}

\subsubsection{Model Selection}

The GitHub issue resolution task is challenging for LLMs, requiring understanding the issue and locating the edition location in the codebase and generating correct patches. Considering the difficulty of this task,  we need to choose LLMs with strong coding capabilities and long text understanding abilities. In addition, since some task instances on \Swebenchx include visual input such as images, we also need to select LLMs with visual understanding capabilities. We select three representative state-of-the-art LLMs for evaluation: \GPTFullName~\cite{gpt-4o}, \ClaudeFullName~\cite{claude-3-5-sonnet} and \DeepSeekFullName~\cite{deepseekV25}, as shown in Table~\ref{table:models}. For simplicity, we refer to these models as \GPT, \Claude, and \DeepSeek in the following sections. 

\begin{table}[t]
  \centering
  \footnotesize
  \setlength\tabcolsep{6pt}  
     \caption{Overview of evaluated LLMs.}
     \tabmargin
  \renewcommand{\arraystretch}{1.2}  

  \resizebox{0.95\columnwidth}{!}{
    \begin{tabular}{lccccc}
    \toprule
    \textbf{Model} & \textbf{Company} & \textbf{Size} & \textbf{Context Window} & \textbf{With Visual Ability} & \textbf{Release Date} \\
    \midrule
    \DeepSeekFullName & DeepSeek & 236B & 128K & No & Sep 2024 \\
    \GPTFullName & OpenAI & Unknown & 128K & Yes & Aug 2024 \\
    \ClaudeFullName  & Anthropic & Unknown & 200K & Yes & Jun 2024 \\
    \toprule
    \end{tabular}
  }

      \label{table:models}
\end{table}

\subsubsection{Evaluation Method}  
Due to the complexity of issue resolution tasks, it is challenging to obtain correct results by directly feeding issue descriptions and codebases into LLMs. Therefore, we evaluate LLMs using some advanced GitHub issue resolution approaches. As mentioned in Section~\ref{sec:methodBackground}, we can divide current issue resolution approaches into three types following previous studies~\cite{anthropic2024building,ouyang2024repograph,pan2024training}: retrieval augmentation-based method, LLM workflow-based method, and LLM agent-based method. To ensure a comprehensive evaluation, we select one representative method from each category. Based on the SWE-bench leaderboard~\cite{swebench2024leaderboard}, we choose baselines with high performance and reasonable cost. Given the limited performance of current RAG-based methods~\cite{tao2024magis,jimenez2023swe}, we choose the oracle retrieval method for evaluation. For the LLM workflow-based method, we select the Agentless~\cite{xia2024agentless} method, which is ranked 5th in the SWE-bench-Verified~\cite{swebench2024leaderboard}. And for the agent-based method, we choose the AutoCoderOver~\cite{zhang2024autocoderover} method, which is ranked 2nd in the SWE-bench-Full~\cite{swebench2024leaderboard}. These methods not only demonstrate advanced performance but also have reasonable costs according to previous studies~\cite{xia2024agentless}.

Finally, we use three methods: the oracle retrieval~\cite{jimenez2023swe,tao2024magis}, the Agentless method~\cite{xia2024agentless} and the AutoCodeRover method~\cite{zhang2024autocoderover}. The descriptions of these methods are listed as follows:

\textbf{Oracle Retrieval.} The oracle retrieval~\cite{tao2024magis,jimenez2023swe} method provides both the issue description and the specific code files that require editing as inputs. These files are identified directly from the gold patch in each task instance. This approach enables an ideal-condition evaluation, simulating a scenario where developers accurately locate the files necessary for resolving the issue.


\textbf{Agentless-X.} 
Considering that current issue resolution frameworks are designed primarily for Python, we select the Agentless approach~\cite{xia2024agentless} as a baseline and adapt it to support additional languages (e.g., JavaScript, TypeScript, and Java), enabling evaluation on \Swebenchx. We refer to this multilingual adaptation as Agentless-X. We choose Agentless as our baseline because it can achieve comparable performance while maintaining lower costs compared to other methods.

The original Agentless approach is a two-phase method for issue resolution. First, the localization phase uses a hierarchical approach where the LLM successively identifies the relevant files, specific classes or methods, and finally, locates the exact code lines to be edited. Second, in the  repair phase, the LLM generates a patch based on the locations identified in the first phase.

In Agentless-X, we retain Agentless's original design and expand it to support multiple languages. The original Agentless approach uses AST tools to parse file structures during the localization stage. In Agentless-X, we use different AST tools for each language: Babel~\cite{babel} for JavaScript/TypeScript and JParser~\cite{jparser} for Java. Additionally, we rewrite LLM prompts in multilingual versions. For parameter settings, we follow the default setting from the original Agentless implementation.\footnote{\url{https://github.com/OpenAutoCoder/Agentless}. We use the Agentless-v1 for evaluation.}

\textbf{AutoCodeRover-X.\footnote{We use the AutoCodeRover(v20240620) for evaluation.}} We choose the AutoCodeRover~\cite{zhang2024autocoderover} 
method as our baseline. Similar to the Agentless method, the AutoCodeRover method only supports Python language. We adapt this method to support additional languages (e.g., JavaScript, TypeScript, and Java), enabling evaluation of our multilingual benchmark. We refer to this multilingual adaption as AutoCodeRover-X.

The original AutoCodeRover method resolves issues using two LLM-based agents: the context retrieval agent and the patch generation agent. First, given issue descriptions and codebase as input, the context retrieval agent searches code information related to the given issue. In this stage, the agent searches relevant code information by calling code search tools iteratively until the agent thinks that enough code information is retrieved. Then, the retrieved code information is input to the patch generation agent. This agent will keep refining the generated patch until it can be applied successfully or until the maximum number of attempts is reached.

In AutoCodeRover-X, we retain AutoCodeRover's original design and expand it to support multiple languages. The context retrieval agent uses AST tools to parse file structures. In AutoCodeRover-X, we use different AST tools for each language: Babel~\cite{babel} for JavaScript/TypeScript and tree-sitter~\cite{tree-sitter} for Java. Additionally, we rewrite LLM prompts in multilingual versions. For parameter settings, we follow the default setting from the original AutoCodeRover implementation~\cite{auto-coderover-config}.



\subsubsection{Evaluation Metrics}
Following previous studies~\cite{jimenez2023swe,tao2024magis,xia2024agentless}, we use these metrics to evaluate the performance of LLMs:
\begin{itemize}[left=10pt]
    \item \textbf{Resolve Rate:} the percentage of task instances that are resolved. 
    \item \textbf{Apply Rate:} the percentage of generated patches that are intergrated to codebase using Git~\cite{git} tool without any errors. 
    \item \textbf{Cost:} the average cost of evaluating task instances.
\end{itemize}

\subsection{Performance of LLMs on \Swebenchx}

We evaluate three advanced LLMs with three methods on \Swebenchx.  The results are presented in Table~\ref{table:performanceAll}. First, we can find that performance of advanced LLMs on \Swebenchx remains limited. Among the evaluated models, \GPT with Agentless-X method achieves the highest resolve rate of 8.6\% and apply rate of 87.4\% on \Swebenchx. 
And \GPT with AutoCodeRover-X method achieves the second best performance with the resolve rate of 8.1\% and apply rate of 85.8\% on \Swebenchx. 
 Moreover, \Claude with oracle retrieval method achieves the third best performance with the resolve rate of 7.8\% and apply rate of 61.8\% on \Swebenchx.
However, the performance of these LLMs across both methods remains limited,  highlighting the need for further improvements in the issue resolution ability of LLMs.

Second, the results show that the Agentless-X method and the AutoCodeRover-X outperform the oracle retrieval method for both \GPT and \DeepSeek. Specifically, for \GPT, Agentless-X increases the resolve rate from 2.5\% to 8.6\%, while AutoCodeRover-X improves it to 8.1\%. 
Similarly, \DeepSeek's performance also improves, with Agentless-X raising the resolve rate from 2.7\% to 3.9\% and AutoCodeRover-X enhancing it further to 6.0\%. These results demonstrate the effectiveness of the Agentless and the AutoCoderRover in improving the issue resolution ability of LLMs.


Third, we find that for \Claude, both the oracle retrieval method and the AutoCodeRover-X method perform better, achieving resolve rates of 7.8\% and 7.6\%, respectively, while its performance with the Agentless-X method is limited to only 1.9\%. 
 We believe this higher performance with oracle retrieval is due to \Claude's strong capability in understanding long code segments, which allows it to make accurate code changes when the correct file locations are provided. In contrast, with Agentless-X method, \Claude's performance is limited because it often fails to produce results in the expected format during the localization stage. This inconsistency makes it hard to parse the location output correctly, ultimately preventing patch generation. The detailed discussion of this issue is in Section~\ref{sec:parseError}. Compared to the Agentless-X method, AutoCodeRover-X achieves better performance on \Claude, demonstrating its robustness in enhancing the performance of different LLMs.

Last, we observe that both the Agentless-X method and the AutoCodeRover-X method perform worse on JavaScript and TypeScript than on Java and Python. This is likely due to the design of these methods, which both rely on AST tools to parse repository files and locate key structures, such as classes and methods. Specifically, they focus on extracting information from traditional object-oriented constructs like classes and methods but do not account for other critical structures.  In Python and Java, essential information is typically organized within classes and methods, so this AST-based extraction is effective. However, in JavaScript and TypeScript repositories, code frequently relies on types and interfaces to define important structures that are not part of the traditional class or method constructs. This makes it challenging for the original extraction method to capture this information effectively. Additionally, JavaScript and TypeScript always use anonymous and arrow functions, which AST tools struggle to parse compared to standard functions. For these methods to handle JavaScript and TypeScript more effectively, it would need to account for these language-specific features, such as arrow functions, anonymous functions, and the use of types and interfaces.


\begin{table}[t]
  \centering
  \small 
  \setlength\tabcolsep{3pt}
  \caption{Performance comparison across different LLMs on \Swebenchx. In the table, DeepSeek-V2.5 is short for DeepSeek, GPT-4o-2024-08-06 is short for GPT-4o, and Claude-3.5-Sonnet-2024-06-25 is short for Claude-3.5.}
  \tabmargin
  \label{table:performanceAll}
  \resizebox{\columnwidth}{!}{
    \begin{tabular}{cccccccccc}
    \toprule
    \multirow{2}{*}{\textbf{Model}} & \multicolumn{3}{c}{\textbf{Oracle Retrieval}} & \multicolumn{3}{c}{\textbf{Agentless-X}} & \multicolumn{3}{c}{
    {\textbf{AutoCodeRover-X}}} \\
    \cmidrule(r){2-4} \cmidrule(r){5-7} \cmidrule(r){8-10}
    & \textbf{Resolve Rate} & \textbf{Apply Rate} & \textbf{Cost(\$)} & \textbf{Resolve Rate} & \textbf{Apply Rate} & \textbf{Cost(\$)} & \textbf{Resolve Rate} & \textbf{Apply Rate} & \textbf{Cost(\$)} \\
    \midrule
    
    \textbf{DeepSeek (All)} & 2.7\% (26/959)  & 49.0\% (470/959) &  
    0.005& 3.9\% (37/959) & 52.0\% (499/959) & 0.010 & 6.0 \% (58/959) & 53.0\% (508/959) & 0.028  \\
    \midrule
    \quad Python & 1.3\% (5/374) & 49.7\% (186/374)& 0.008 & 6.1\% (23/374)& 70.9\% (265/374)& 0.014 & 7.2 \% (27/374) & 53.2\% (199/374) & 0.026 \\
    \quad TypeScript & 1.9\% (4/210)& 45.2\% (95/210)& 0.003 & 1.9\% (4/210)& 44.3\% (93/210)& 0.006  & 4.3 \% (9/210) & 46.7\% (98/210) & 0.033\\
    \quad JavaScript & 2.2\% (6/270)& 48.5\% (131/270)& 0.003 & 2.6\% (7/270)& 37.0\% (100/270) & 0.009 &3.7 \% (10/270) & 53.3\% (144/270) & 0.029 \\
    \quad Java & 10.5\%  (11/105)& 55.2\% (58/105) & 0.038 & 2.9\%  (3/105)&39.0\% (41/105)& 0.007 &11.4 \% (12/105) & 63.8\% (67/105) & 0.021\\
    \midrule
    
    \textbf{\GPT (All)} & 2.7\% (26/959)  & 42.6\% (409/959) & 0.069 & \textbf{8.6\% (82/959)} & \textbf{87.4\% (838/959)} & 0.098 &8.1 \% (78/959) & 85.8\% (823/959) & 0.168\\
    \midrule
    \quad Python &1.9\% (7/374) &42.5\% (159/374) & 0.112 & 8.8\% (33/374) & 89.3\% (334/374) & 0.090 &9.9 \% (37/374) & 89.6\% (335/374) & 0.156\\
    \quad TypeScript & 3.3\% (7/210) &44.3\% (93/210) & 0.035 &6.2\% (13/210) &85.2\% (179/210) & 0.095 &6.2 \% (13/210) & 78.6\% (165/210) & 0.179\\
    \quad JavaScript &1.9\% (5/270) & 40.0\% (108/270) & 0.049 &6.3\%  (17/270)&87.8\% (237/270) & 0.113 &3.7 \% (10/270) & 84.4\% (228/270) & 0.193\\
    \quad Java &6.7\% (7/105)& 46.7\%  (49/105) & 0.003 & 18.1\% (19/105) &83.8\% (88/105) & 0.053 &17.1 \% (18/105) & 90.5\% (95/105) & 0.119\\
    \midrule                                                 
    
    \textbf{Claude-3.5 (All)} &7.8\% (75/959) & 61.8\% (593/959) & 0.110 & 1.9\% 
 (18/959)& 14.6\%(140/959)& 0.272 &7.6 \% (73/959) & 61.5\% (590/959) & 0.325 \\
    \midrule
    \quad Python & 5.1\% (19/374) & 52.9\% (198/374) & 0.179 & 1.6\%  (6/374)& 1.6\% (55/374) &0.335 &8.8 \% (33/374) & 80.5\% (301/374) & 0.260\\
    \quad TypeScript & 6.7\% (14/210) & 70.0\% (147/210) & 0.047 & 1.0\% (2/210) & 10.5\% 
 (22/210)& 0.170 &4.3 \% (9/210) & 33.8\% (71/210) & 0.353 \\
    \quad JavaScript & 8.5\%  (23/270)& 63.3\% (171/270) & 0.073 & 2.2\% (6/270)& 18.5\% (50/270) & 0.178 &4.1 \% (11/270) & 46.7\% (126/270) & 0.392\\
    \quad Java & 18.1\% (19/105) & 73.3\% (77/105) & 0.063 &3.8\% (4/105|)& 12.4\% (13/105) & 0.342 &19.0 \% (20/105) & 87.6\% (92/105) & 0.327\\
    \bottomrule
    \end{tabular}
    }

\end{table}
\tabmargin

\begin{center} 
\begin{myboxc} \textbf{RQ1 Summary:} The performances of current LLMs remain limited on \Swebenchx. Besides, the Agentless-X method and the AutoCodeRover-X can effectively enhance issue resolution preformance of \GPT and \DeepSeek. However, while \Claude achieves good performance with the oracle retrieval method and the AutoCodeRover-X method, it struggles with Agentless-X. Lastly, Agentless-X and AutoCodeRover-X perform better on Java and Python compared to JavaScript and TypeScript. \end{myboxc} 
\end{center}

\subsection{Performances of LLMs on Issues With Visual Inputs }

We evaluate two advanced LLMs with visual understanding capabilities, \GPT and \Claude, on a set of 19 task instances that require understanding images in issues. 
This evaluation uses all three baselines. Additionally, given the relatively small size of the subset, we repeat each experiment three times to minimize the effects of randomness. To investigate the effect of visual inputs, we conduct experiments under three different settings: 

\begin{itemize}[left=10pt]
     \item \textbf{Text Only.} In this setting, we do not provide the actual images from issues to LLMs. Instead, we only include the image URLs as text, without any visual information.

    \item \textbf{Text \& Image.} In this setting, both the text and images from issues are input to LLMs. The LLMs use its visual capabilities to understand the images and resolve the issue.

 \item \textbf{Image-augmented Text.} This is a two-stage approach. In the first stage, we use a prompt to instruct models to understand images and to rewrite the issue text with visual information. In the second stage, the rewritten text, augmented with image details, is used as input to the model. Compared to the second setting, this approach allows us to better understand how the model processes and interprets the visual information.

\end{itemize}

 The results are shown in Table~\ref{table:performanceVisual}, highlighting three key conclusions. First, current LLMs show limited performance on issues requiring image understanding. Using oracle retrieval method, which provides the correct files, \Claude achieves the best performance, resolving only 10.5\% of the issues in the image-augmented text setting. 
In contrast, \GPT resolves only 5.3\% of issues with the AutoCodeRover-X method. Second, leveraging visual information is beneficial for resolving issues requiring image understanding. When using oracle retrieval method, compared to the ``Text only'' setting, \Claude's resolve rate increases from 3.5\% to 10.5\% in the ``Image-augmented Text'' setting. Similarly, with the AutoCodeRover-X method, \GPT does not resolve any issue in the ``Text only'' setting. However, when image information is included, the resolve rate increases to 3.1\% in the ``Text \& Image'' setting and further improves to 5.3\% in the ``Image-augmented Text'' setting. Third, LLMs using the Agentless-X method and the AutoCodeRover-X method both struggle to resolve issues requiring image understanding.  Despite its solid performance on \Swebenchx, Agentless-X failed to resolve any issues requiring image understanding on both models. Similarly, AutoCodeRover-X also shows limited performance in this area, with \GPT resolving at most 5.3\% of issues, while \Claude failed to resolve any. These results highlight that current methods need improvements to utilize visual information effectively.

\begin{table}[t]
  \centering
  \setlength{\tabcolsep}{2pt}
    \caption{Performance comparison across different LLMs on \Swebenchx-V. In this table, Claude-3.5-Sonnet-2024-06-25 is short for Claude-3.5 and the ``Image-augmented Text'' setting is short for IAG-Text. }
  \label{table:performanceVisual}
  \tabmargin
  \resizebox{\columnwidth}{!}{
    \begin{tabular}{llccccccccc}
    \toprule
    \multirow{2}{*}{\textbf{Model}} & \multirow{2}{*}{\textbf{Setting}} & \multicolumn{3}{c}{\textbf{Oracle Retrieval}} & \multicolumn{3}{c}{\textbf{Agentless-X}} & \multicolumn{3}{c}{\textbf{AutoCodeRover-X}}\\
    \cmidrule(r){3-5} \cmidrule(r){6-8}\cmidrule(r){9-11}
    & & \textbf{Resolve Rate} & \textbf{Apply Rate} & \textbf{Cost(\$)} & \textbf{Resolve Rate} & \textbf{Apply Rate} & \textbf{Cost(\$)} & \textbf{Resolve Rate} & \textbf{Apply Rate} & \textbf{Cost(\$)} \\
    \midrule
    \multirow{3}{*}{\GPT} 
    & Text Only & 1.6\% (1/57) & 45.6\% (26/57) & 0.036 & 0\% (0/57) & 42.1\% (24/57) & 0.096 & 0\% (0/57) & 84.2\% (48/57) & 0.159\\
    & Text \& Images & 1.6\% (1/57) & 33.3\% (19/57)  & 0.038 & 0\% (0/57) & 26.3\% (15/57) & 0.110 & 3.5\% (2/57) & 77.2\% (44/57) & 0.220\\
    & IAG-Text & 1.6\% (1/57) & 52.6\% (30/57) & 0.046 & 0\% (0/57) & 38.6\% (22/57) & 0.109 & 5.3\% (3/57) & 78.9\% (45/57) & 0.269\\
    \midrule
    \multirow{3}{*}{Claude-3.5} 
    & Text Only & 3.5\% (2/57) & 50.9\% (29/57) & 0.058 & 0\% (0/57) & 57.9\% (33/57) &  0.210 & 0\% (0/57) & 84.2\% (48/57) & 0.271\\
    & Text \& Images & 3.5\% (2/57) & 59.6\% (34/57) & 0.062 & 0\% (0/57) & 63.2\% (36/57) &  0.224 & 0\% (0/57) & 61.4\% (35/57) & 0.309\\
    & IAG-Text & 10.5\% (6/57) &  59.6\% (34/57) & 0.074 & 0\% (0/57) & 52.6\% (30/57) & 0.233 & 0\% (0/57) & 68.4\% (39/57) & 0.339\\
    \bottomrule
    \end{tabular}
  }

\end{table}

\begin{center}
    \begin{myboxc} \textbf{RQ2 Summary: } 
     Our evaluation shows that current LLMs struggle with issues requiring image understanding. Besides,  leveraging visual information is beneficial for resolving issues requiring image understanding. However, we find LLMs with Agentless-X and AutoCodeRover-X show limited performances on issues requiring image understanding, highlighting the need for these methods to utilize visual information  effectively.
    \end{myboxc} 
\end{center}


   


\subsection{Failure Analysis}

To investigate why LLMs fail to resolve issues on \Swebenchx, we analyze their behavior at the localization and patch generation stages, identifying one type of failure at each stage.

\begin{figure}[t]
    \centering
    \includegraphics[width=0.95\linewidth]{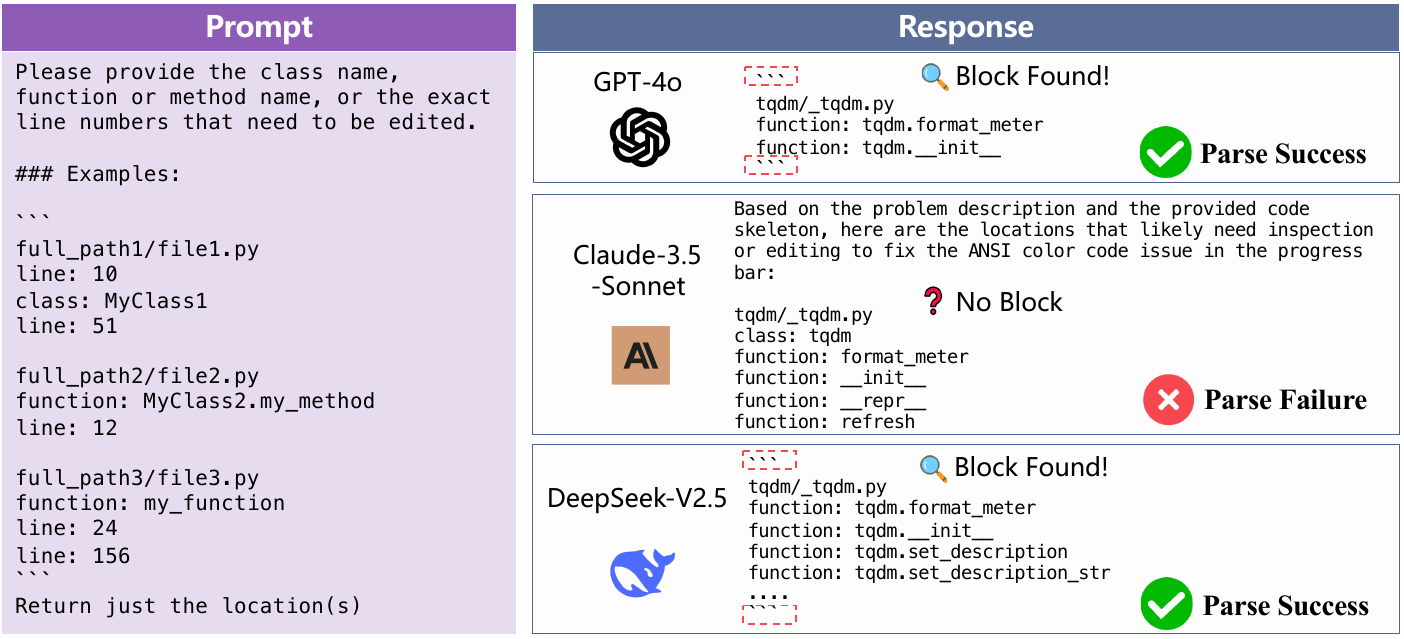}
    \figmargin
    \caption{Responses of different LLMs in the localization stage of Agentless-X. }
    \label{fig:ErrorFormat}
\end{figure}

\textbf{Parsing failure of structural output.} \label{sec:parseError}
This type of failure occurs in the localization stage of the Agentless-X method. It happens when the model fails to follow the specified output format given in the prompt.
In the example shown in Figure~\ref{fig:ErrorFormat}, the prompt instructs LLMs to provide the locations to be edited (e.g., class names, function names, method names, and line numbers) and to wrap these results with a code block. Then, this method uses regex expressions to parse the localization results from responses of LLMs. We can find that both \GPT and \DeepSeek follow the prompt's instructions, successfully enclosing their output in code blocks, allowing the results to be parsed correctly and used in the following repair stage. However, \Claude does not follow this format; instead of enclosing the location information in a code block, it provides the output in plain text, which leads to a parsing failure. As a result, no location information is passed to the repair stage, and no patch can be generated from \Claude in the following repair stage.  

To further explore the impact of this error, we compute the parsing success rate for each model in the localization stage of the Agentless-X framework, as shown in Table~\ref{table:ErrorFormatTable}. \GPT achieves the highest parsing success rate at 97.2\%, while \Claude obtains a much lower success rate of 11.8\%. This parsing error significantly affects \Claude's ability to generate patches, leading to a lower overall resolve rate (1.9\%). However, we find that this failure can be mitigated effectively by adding format constraints in original prompt. In our experiments, we append the original prompt with ``You must wrap the results with \verb|```|''. We find this prompt increases the \Claude's parsing success rate and hence increases the resolve rate from 1.9\% to 7.4\%. This results highlight the importance of building robust prompts in current methods.

\begin{table}[t]
  \centering
  \small 
   \caption{Parsing success rate of LLMs with the Agentless-X method on \Swebenchx.}
  \label{table:ErrorFormatTable}
  \tabmargin
    \begin{tabular}{lcccc}
    \toprule
    \multirow{2}{*}{\textbf{Model}} & \multicolumn{4}{c}{\textbf{Agentless-X}}  \\
    \cmidrule(r){2-5}
   
    &\textbf{Prompt Setting} & \textbf{Parsing success Rate} & \textbf{Resolve Rate} & \textbf{Apply Rate}  \\
   \midrule

    \GPT & default &97.2\% (932/959) &8.6\% (82/959) & 87.4\% (838/959)  \\
    \DeepSeek & default& 82.5\% (791/959)  & 3.9\% (37/959) & 52.2\% (501/959)  \\
    \Claude & default & 10.0\% (96/959) &   1.9\% (18/959) &  17.6\% (169/959)   \\
    \Claude & new prompt & 98.5\% (945/959) &  7.4\% (71/959)  & 63.3\% (607/959) \\  
     
    \bottomrule
    \end{tabular}

\end{table}

\textbf{Insufficient ability of cross-file issue resolution.} This type of failure occurs in the patch generation stage of all evaluated baselines. In the issue resolution task, resolving issues that require modifications across multiple files is particularly challenging. To investigate LLMs' ability to resolve single-file and cross-file issues, we categorize issues on \Swebenchx into these two types.  For each issue, we use its gold patch as a reference and check how many files need to be modified to resolve the issue. If multiple files are modified in the gold patch of the issue, we classify this issue as a cross-file issue; otherwise, it is categorized as a single-file issue. And then we analyze performances of LLMs in two types of issues. As shown in Table~\ref{table:CrossFileEval}, we can observe that LLMs perform significantly worse on cross-file issues compared to single-file issues.  

We assume one reason LLMs perform poorly on cross-file issues is their tendency to modify only a single file. To investigate this assumption, we first examine the number of files modified in the patches generated by LLMs for cross-file issues. Then, we calculate the proportion of patches that include modifications to multiple files versus those with changes to only a single file. We find that, even with the oracle retrieval method, which provides the relevant files to the model, LLMs still primarily edit a single file. Specifically, \Claude modifies a single file in 74.9\% of issues, \GPT in 72.9\%, and \DeepSeek in 86.3\%. Moreover, with the Agentless-X method, all models modify a single file in 100\% of issues. Additionally, with the AutoCodeRover-X method, \Claude modifies a single file in 93.3\% of issues, \GPT in 82.4\%, and \DeepSeek in 90.0\%.  These results highlight the insufficient ability of current LLMs in resolving issues that require modifying multiple files.




\begin{table}[t]
  \centering
  \small 
  \setlength\tabcolsep{4pt}
      \caption{Performance comparison across different LLMs on \Swebenchx for cross-file and single-file issues.}
  \label{table:CrossFileEval}
  \tabmargin
  \resizebox{0.98\columnwidth}{!}{
    \begin{tabular}{llcccc}
    \toprule
    \multirow{2}{*}{\textbf{Model}} & \multirow{2}{*}{\textbf{Method}} & \multicolumn{2}{c}{\textbf{Single-file Issues}} & \multicolumn{2}{c}{\textbf{Cross-file Issues}} \\
    \cmidrule(r){3-4} \cmidrule(r){5-6}
    & & \textbf{Resolve Rate} & \textbf{Apply Rate} & \textbf{Resolve Rate} & \textbf{Apply Rate} \\
    \midrule
    \multirow{3}{*}{\GPT} 
    & Oracle Retrieval & 3.5\% (21/601) & 42.9\% (258/601) & 1.4\% (5/358) & 42.2\% (151/358) \\
    & Agentless-X &  12.0\% (72/601) & 87.4\% (525/601) & 2.8\% (10/358)  & 87.4\% (313/358) \\
       & AutoCodeRover-X &  11.5\% (69/601) & 88.4\% (531/601) & 2.5\% (9/358)  & 81.6\% (292/358) \\

    \midrule
    \multirow{2}{*}{\DeepSeek} 
    & Oracle Retrieval & 3.7\% (22/601)  & 50.6\% (304/601) & 1.1\% (4/358)  & 46.4\% (166/358) \\
    & Agentless-X & 5.2\% (31/601)  & 52.4\% (315/601) & 1.7\% (6/358)  & 51.4\% (184/358) \\
    & AutoCodeRover-X &  8.5\% (51/601) & 55.9\% (336/601) & 1.9\% (7/358)  & 48.0\% (172/358)  \\

    \midrule
    \multirow{2}{*}{\Claude} 
        & Oracle Retrieval & 10.5\% (63/601) & 66.1\% (397/601) & 3.4\% (12/358) & 54.7\% (196/358) \\
    & Agentless-X & 2.5\% (15/601) & 13.6\% (82/601) & 0.8\% (3/358) & 16.2\% (58/358) \\
    & AutoCodeRover-X &  10.8\% (65/601) & 63.4\% (381/601) & 2.2\% (8/358)  & 58.3\% (209/358)  \\
    \bottomrule
    \end{tabular}
    }

\end{table}


\begin{center}
    \begin{myboxc} \textbf{RQ3 Summary: } Our analysis shows that \Claude struggles to generate results in the expected format during the Agentless-X localization stage, leading to poor issue resolution performance. We find that adding specific format instructions to the prompt can improve \Claude's results. Additionally, current LLMs perform poorly on cross-file issues and tend to modify only a single file, even when multiple files require changes.
    \end{myboxc} 
\end{center}


\secmargin
\section{Discussion}
\hspace{1em}\textbf{Potential Impacts.} First, we propose a GitHub issue resolution benchmark that captures diversity across programming languages, repository domains, and modalities of input information. This allows researchers to more effectively evaluate any method of resolving issues in different programming languages and with multimodal data. Second, our error analysis highlights specific limitations in current methods and LLMs, such as poor performance on TypeScript and JavaScript issues and challenges in resolving cross-file issues, etc. These findings provide insights for improving LLMs and issue resolution techniques. Third, we have made our data collection code and tutorial publicly available, allowing other researchers to collect GitHub issue resolution data to build a new evaluation benchmark or training dataset.  

\textbf{Limitations and future directions.} First, following the SWE-bench~\cite{jimenez2023swe}, we collect data based on repository popularity, which may introduce bias because of overlooking less popular but potentially valuable repositories. To address this, we have open-sourced our data collection code and tutorial, enabling other researchers to collect data from repositories they consider valuable. Additionally, we plan to expand our benchmark in the future to include data from a broader range of repositories to mitigate this bias. Second, following the SWE-bench~\cite{jimenez2023swe}, we use attribute filtering and execution-based filtering to remove invalid instances. However, this approach may filter out some instances. For example, using path keywords to identify test files could miss files that don't follow conventional naming patterns, and filtering out instances without FAIL2PASS tests excludes issues like performance improvements. We do this to ensure that the tests for each instance can reliably evaluate the correctness of the solution. In the future, we plan to use more advanced techniques, like machine learning-based file identification, to identify test files. Besides, we plan to expand the benchmark to include more issue types. Finally, we do not evaluate the effects of web information, as there are currently no methods to resolve issues using the website information contained in the issue description. In the future, we will evaluate potential approaches to incorporate web-based information into issue resolution.

\secmargin

\section{Conclusion}
In this paper, we propose \Swebenchx, a GitHub issue resolution benchmark with multi-aspect
diversity in programming languages, repository domains and modality of input information. The evaluation results demonstrate that current LLMs show limited performances on \Swebenchx. Besides, we also find that current LLMs struggle to resolve issues that require understanding images. Furthermore, we analyze the reasons behind LLM's failure on \Swebenchx, to shed light on future performance improvements of LLMs on \Swebenchx.

\secmargin
\section{Data Availability}
\label{sec:open-science}
Our code and  data are available at \url{https://github.com/DeepSoftwareAnalytics/OmniGIRL}.

\sloppy
\begin{acks}
This work is supported by CCF-Huawei Populus Grove Fund CCF-HuaweiSE202403, the National Natural Science Foundation of China (62032025) and the Guangdong Basic and Applied Basic Research Foundation (2023A1515012292).
\end{acks}


\bibliographystyle{ACM-Reference-Format}
\bibliography{ref}

\end{document}